\documentclass[12pt]{iopart}
\usepackage{amssymb}
\usepackage{txfonts}

\usepackage{iopams}
\usepackage{graphics}
\usepackage{color} 
\usepackage{CJK}
\newcommand{\beq}{\begin{equation}}
\newcommand{\eeq}{\end{equation}}
\newcommand{\beqa}{\begin{eqnarray}}
\newcommand{\eeqa}{\end{eqnarray}}

\begin{document}

\title[Electronic analogy of Goos-H\"{a}nchen effect: a review]{Electronic analogy of Goos-H\"{a}nchen effect: a review}

\author{Xi Chen$^{1,2}$, Xiao-Jing Lu$^{1}$, Yue Ban$^{2}$, and Chun-Fang Li$^{1}$}

\address{$^{1}$ Department of Physics, Shanghai University, 200444 Shanghai, China}
\address{$^{2}$ Departamento de Qu\'{\i}mica-F\'{\i}sica, UPV-EHU, Apdo 644, 48080 Bilbao, Spain}

\ead{xchen@shu.edu.cn}

\begin{abstract}
The analogies between optical and electronic Goos-H\"{a}nchen effects are established based on electron wave optics in semiconductor or graphene-based nanostructures.
In this paper, we give a brief overview of the progress achieved so far in the field of electronic Goos-H\"{a}nchen shifts, and show
the relevant optical analogies. In particular, we present several theoretical results on the giant positive and negative Goos-H\"{a}nchen shifts
in various semiconductor or graphen-based nanostructures, their
controllability, and potential applications in electronic devices, e.g. spin (or valley) beam splitters.

\end{abstract}


\submitto{\JOA}
\maketitle

\section{ Introduction}

The Goos-H\"{a}nchen (GH) effect, named after Hermann Fritz Gustav Goos
and Hilda H\"{a}nchen \cite{Goos}, is an optical phenomenon in which
a light beam undergoes a lateral shift from the position predicted
by geometrical optics, when totally reflected from a
single interface of two media having different refraction indices
\cite{Lotsch}. The lateral GH shift, conjectured by Isaac Newton in the 18th century \cite{Newton},
was theoretically explained by Artmann's stationary phase
method \cite{Artmann} and Renard's energy flux method \cite{Renard}.
With the development of laser beam and integrated optics \cite{Lotsch}, the GH shift becomes very significant nowadays,
e.g. in the applications of optical waveguide switch \cite{Sakata} and sensors
\cite{Yin-Hesselink,Yu}, or in fundamental problems on tunneling
time \cite{Steinberg-C,Stahlhofen,Balcou}. Therefore, the GH shift as well as
other three nonspecular effects such as angular
deflection, focal shift, and waist-width modification have been investigated in various configurations like frustrated total internal reflection
(FTIR) \cite{Ghatak,Haibel,Chen-PRAa,Chen-OL}, attenuated total
reflection (ATR) \cite{Yin,Pillon}, and partial reflection
\cite{Hsue-T,Riesz,Li-2,Nimtz,Chen-PRAb,Chen-JOA}, and also extended to other physics fields
\cite{Lotsch} including nonlinear optics, plasmas, semiconductor,
acoustics \cite{Briers}, neutron physics \cite{Ignatovich,Haan} and
even atom optics \cite{Zhang-WP}.

The considerable number of publications on the subject, and a recent
workshop on ``Beam Shifts: Analogies between light and matter waves"
held at Lorentz Center, Leiden, Netherlands (28 March-1 April
2011) demonstrate much current interest, not only from optics
community but also from other fields such as condensed matter and
particle physics. In recent years, the lateral GH shift and
transverse Imbert-Fedorov (IF) shift (also known as spin-Hall effect of light)
\cite{Fedorov,Imbert,Costa,Li-PRA,Hosten,Aiello}, have
attracted much attention (see review \cite{reviewoptics}).
The field covered here is vast, and we pay special attention to work done by the authors,
while making effort to offer a global perspective. Therefore, in
this review we shall concentrate on the electronic GH shifts in
semiconductor and graphene-based nanostructures and their
applications in spin (or valley) beam splitters and filters.

Historically, the study of electronic GH shift is traced back to
1960s \cite{Renard,Carter}. In 1964, Renard \cite{Renard} proposed
the energy flux method to investigate the GH shift for matter waves,
based on the analogy between Schr\"{o}dinger equation in quantum
mechanics and Helmholtz equation in electromagnetic theory. Later
on, Carter and Hora \cite{Carter} considered the GH shift of matter
waves in total reflection at grazing incidence. Motivated by the
progress on optical GH and IF shifts,  Miller \textit{et. al.}
\cite{Miller} made a theoretical investigation of the analogous GH
shifts of a Dirac electron beam which undergoes a series of total
internal reflections from finite potential barriers of arbitrary
smoothness. In general, it is found that the GH shift for Dirac
electrons is at most of the order of a Compton wavelength. To
observe it macroscopically, the GH shift for Dirac electrons
was further proposed to be amplified by multiple reflections
\cite{Fradkin}. However, it is still an open challenge to detect the electric GH shift,
due to the smallness of GH shift, and the difficulty in preparation of well-collimated electron beam,
the analogue of an optical beam in electronics.

With the advent of techniques for the semiconductor growth and
fabrication of semiconductor nanoelectronic devices, the optics-like
phenomenon \cite{Gaylord-IEEE}, such as reflection, focusing,
diffraction, and interference, resulting from the fact that the
quantum-mechanical wave nature of electrons, become prominent in
electron wave devices, thus have given rise to a field of research
which is best described as ballistic electron optics in
two-dimensional electron systems (2DESs) \cite{Datta,Dragoman-D}. In
1993, Wilson \textit{et. al.} \cite{Wilson-G-G} have studied the GH
shift and associated time delay for an electron beam totally
reflected from a potential-energy/effective-mass interface in
semiconductor. It was found that the phase shift resulting in the GH
shift and time delay upon the total internal reflection has great
effect on the quantum interference in electron wave devices, and
thus contributes to novel electron waveguiding characteristics.
Again, the smallness of electronic GH shifts in total reflection has
impeded its direct measurement and applications in electron devices.
To overcome this bottleneck, we generalized the electronic GH
shift in total internal reflection to that in partial reflection
\cite{ChenPLA06,ChenJAP09}. The resonance-enhanced GH shift of
ballistic electrons in transmission was found in semiconductor quantum barrier
\cite{ChenPLA06} as well as quantum well \cite{ChenJAP09}. More interestingly, the
lateral shifts of electron beam transmitted through a semiconductor quantum well,
acting as electron wave slab, can be negative and positive,
depending on the incidence angle \cite{ChenJAP09}. Moreover, the
negative and positive GH shifts separate the spin-up electron beam
from spin-down electron beam spatially in a more realistic setup, i. e.,
parabolic potential well under a uniform vertical magnetic field
\cite{ChenPRB11} and $\delta$-magnetic barrier nanostructures \cite{ChenPRB08}.
This achievement leads to some other interesting
investigations and applications in quantum electronic devices such
as spin filter or spin beam splitter. For instance, the GH
shift in magnetic-electic nanostructures can be utilized to design
a spin beam splitter, when the GH shift can be controlled by
adjusting the electric potential induced by applied voltage and
magnetic field strength of ferromagnetic stripes \cite{ChenPRB08}.
For recent developments on this line of research, we refer the readers to the references
\cite{Lu-PLA,Lu-EPJB,Lu-JMMM,Lu-JAP}.

A different but relevant line is electronic GH shift in graphene \cite{Beenakker-PRL,Zhao,Chen-EPJB,Guo,Chenarxiv,Manish,Wu,Zhai,CaoPhysB}.
The first experimental fabrication of monolayer graphene \cite{Novoselov,Castro}, the graphitic sheet of one-atom thickness, has inspired many interesting and new concepts on Dirac electron optics to design graphene-based electron wave devices \cite{bookchapter}. Motivated by electronic negative refraction, Veselago lens and foucusing in graphene  \cite{Cheianov}, Beenakker and his collaborators \cite{Beenakker-PRL} have pioneered the quantum GH effect at $p$-$n$ interface in graphene, and have shown
that the electronic GH shifts result in a remarkable $8e^2/\hbar$ conductance plateau in $p$-$n$-$p$ graphene junction.
This work stimulates other progress on resonance-enhanced GH shifts in various graphen-based nanostructures,
including single \cite{Chen-EPJB}, (asymmetric) double barrier \cite{Guo},
and superlattice \cite{Chenarxiv}. These huge (negative or positive) GH shifts are considered to be more suitable to design the controllable valley and spin beam splitter.
Similar to those in semiconductor, the GH shifts in graphene can be also modulated by electric and magnetic barriers \cite{Manish}, which
influences the electronic transport in graphene-based electronics. Besides, a different and intriguing phenomenon
is that the GH shifts can be also controlled by using only strain, without requiring any external fields \cite{Wu,Zhai}.
In strained graphene, the lateral shifts, depending on stain tensor and direction \cite{CaoPhysB}, lead to valley-dependent transports and graphene-based valleytronic devices like valley beam splitter \cite{Zhai}.

The review is organized as follows. In Sec. \ref{semiconductor}
and \ref{graphene}, we shall present the work on GH shifts in semiconductor
and graphene nanostructures, respectively. Throughout this review, we attempt to establish analogies between GH shifts of light beams and electron beams in semiconductor or graphene,
and address such issues as (i) the properties of GH shifts in reflection and transmission, (ii) their
controllability, and (iii) their dependence on different degrees of freedom, e. g. spin or valley.
Some remarks and prospects will be presented in Sec. \ref{summary}.

\section{Electronic Goos-H\"{a}nchen shifts in semiconductor}

\label{semiconductor}

\subsection{single interface}

\begin{figure}[]
\begin{center}
\scalebox{0.6}[0.6]
{\includegraphics{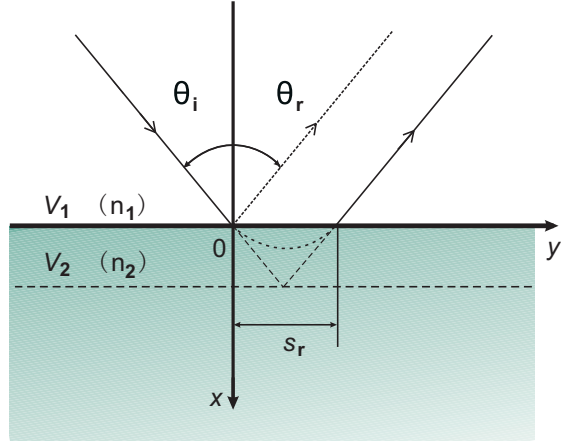}}
\caption{(Color online) Schematic diagram of an electron beam
totally reflected from a single potential energy/effective mass
interface with a lateral GH shift, $s_{r}$. The incident,
reflected are schematically represented by their respective beam
axis. The dotted line is the path expected from the geometrical
result in electron optics. } \label{fig.1}
\end{center}
\end{figure}

We shall start with the GH shift for the totally reflected
electron beam at a single interface between two different potential
energies in semiconductor, which helps to establish the analogy
of optical and electronic GH shifts on a heuristic basis. As shown in Fig. \ref{fig.1},
we consider a ballistic electron beam of total energy $E$ that is incident at an
incidence angle $\theta_0$ from one region to the other, where the
conduction band edge potential energies are  $V_{1}$ and $V_{2}$
($V_{1}<V_{2}$), and the effective mass are $m_{1}^{\ast}$ and
$m_{2}^{\ast}$, respectively. When the incidence angle $\theta_0$ is lager than the
critical angle
\beq
\label{critical angle}
\theta_{c}=\sin^{-1}\sqrt{\frac{m_{2}^{\ast}(E-V_{2})}{m_{1}^{\ast}(E-V_{1})}},
\eeq
the total internal reflection will happen. The critical angle
for total reflection are valid for the potential step (rise), when
the energy satisfies $V_{2}<E<E_{c}$ with
$E_{c}=(m_{2}^{\ast}V_{2}-m_{1}^{\ast}V_{1})/(m_{2}^{\ast}-m_{1}^{\ast})$
\cite{Wilson-G-G}. In this case, the reflectivity and phase shift
are respectively given by
\beq \label{reflectivity} r= \frac{1-i
\frac{m_{1}^{\ast} \kappa}{m_{2}^{\ast} k_{x}}}{1+i
\frac{m_{1}^{\ast} \kappa}{m_{2}^{\ast} k_{x}}} = e^{i \phi_{r}},
\eeq
and
\beq \phi_{r} = -2 \tan{\left(\frac{m_{1}^{\ast}
\kappa}{m_{2}^{\ast} k_{x}}\right)},
\eeq
where $k_{x}=k\cos\theta$,
$k_{y}=k\sin\theta$, $\kappa=(k^{2}_y -k'^{2})^{1/2}$,
$k=[2m_{1}^{\ast}(E-V_{1})]^{1/2}/\hbar$ and
$k'=[2m_{2}^{\ast}(E-V_{2})]^{1/2}/\hbar$. Here $\theta$ is
the incidence angle for the plane-wave component under consideration. Now we
are ready to calculate the GH shift for a finite-sized electron
beam. The wave function of the incident beam is assumed to be
\begin{equation}
\Psi_{in}=\frac{1}{\sqrt{2\pi}}\int_{-\infty}^{\infty}A(k_{y}-k_{y0}) \exp[i(k_{x}x+k_{y}y)]dk_{y},
\end{equation}
with the angular spectrum distribution $A(k_{y}-k_{y0})$ around the central wave vector $k_{y0}$.
The wave function of the reflected one is thus expressed as
\begin{equation}
\Psi_{r}=\frac{1}{\sqrt{2\pi}}\int_{-\infty}^{\infty} r(k_{y}) A(k_{y}-k_{y0}) \exp[i(-k_{x}x+k_{y}y]dk_{y},
\end{equation}
where $r(k_{y})$ is given by Eq. (\ref{reflectivity}).
In practice, the range of the above integral is from $-k$ to $k$,
and here it can be ideally extended from  $-\infty$ to $\infty$ for a well-collimated electron beam.

To find the position where $|\Psi_{r}|^2$ is maximum, that is, the GH
shift of the reflected beam, we look for the position at which the
total phase $\Phi=-k_{x}x+k_{y}y+\phi_{r}$, has an extremum when
differentiated with respected to $k_{y}$, that is to say, $\partial
\Phi / \partial k_{y}=0$ \cite{Bohm}. This provides the stationary-phase expression
of GH shift, namely,
\beq
\label{expressGH}
s_r =-\frac{\partial \phi_{r}}{\partial k_{y}}|_{\theta=\theta_0},
\eeq
which yields the electronic GH shift in total reflection
\cite{Wilson-G-G,ChenPLA06}
\begin{equation}
\label{reflected displacement}
s_{r} = \frac{2}{\kappa_0}\frac{(k^2-k'^2)\tan\theta_0}{(m^{*}_1/m^*_2)\kappa_0^2+(m^{*}_2/m^*_1)
k_{x0}^2},
\end{equation}
where the subscript $0$ denotes the value taken at
$\theta=\theta_{0}$, namely $k_{y} = k_{y0}$.
This demonstrates that there exists the GH shift for ballistic electrons in semiconductor. So far, spin is totally neglected in
our discussions.
Since the GH shift is proportional to the de Broglie wavelength, $1/\kappa_0$, it is a purely wave-like effect, and
has nothing to do with the spin-orbit coupling.
Like the GH shift in optics, the electronic GH shift in total internal reflection is only about
the order of the electron wavelength, which has impeded its direct
measurement and applications in electron devices. This drawback motivates us to study the enhancement of electronic GH shifts,
see below.

\begin{table}[ht]
\caption{Analogy for the electron wave and classical optics} 
\centering 
\begin{tabular}{l|l|l} 
\hline\hline 
Electron wave & TE wave & TM wave \\ [0.5ex] 
\hline 
$ \nabla^2 \Psi = - (2 m/\hbar)(E-V) \Psi$ &  $\nabla^2 E =- \omega^2 \mu \varepsilon E$  &  $\nabla^2 H =- \omega^2 \mu \varepsilon H$ \\ 
$k^2 =2 m(E-V)/\hbar$ ~ (parabolic) &  $k^2 = \omega^2 \mu \varepsilon$ ~ (linear)   & $k^2 = \omega^2 \mu \varepsilon$ ~ (linear) \\ 
$2 (E-V)/\hbar$  & $\omega$ &  $\omega$  \\
$m$ & $\mu$ & $\varepsilon$  \\ 
$1/[2 (E-V)]$ & $\varepsilon$  & $\mu$ \\ [1ex] 
\hline 
\end{tabular}
\label{table1} 
\end{table}

The close analogy between ballistic electrons transport (electron wave propagation) in semiconductor and the electromagnetic wave propagation in classical optics
has been well established \cite{Datta,Dragoman}. The similarity and difference for electron wave and
electromagnetic wave are shown in Table \ref{table1}. As shown in Fig. \ref{fig.1}, we can calculate the GH shift of light beam totally reflected from a single interface
of two media having different refraction indices, $n_1= \sqrt{\mu_1 \varepsilon_1} $ and $n_2= \sqrt{\mu_2 \varepsilon_2} $,
\beqa
s^{TE}_r &=&
\frac{2}{\kappa_0}\frac{(k^2-k'^2)\tan\theta_0}{(\mu_1/\mu_2)\kappa_0^2+(\mu_2/\mu_1)
k_{x0}^2},
\\
s^{TM}_r &=&
\frac{2}{\kappa_0}\frac{(k^2-k'^2)\tan\theta_0}{(\varepsilon_1/\varepsilon_2)\kappa_0^2+(\varepsilon_2/\varepsilon_1)
k_{x0}^2},
\eeqa
which are the  same as Eq. (\ref{reflected displacement}).
With the analog parameters in Table \ref{table1},
we have $k = \sqrt{\mu_1 \varepsilon_1} \omega$ and $k' = \sqrt{\mu_2 \varepsilon_2} \omega$ are the wave vector in two different media,
and other parameters are expressed as $\kappa = (k^2_y - k'^2)^{1/2}$, $k_x = k \cos \theta$, and $k_y = k \sin \theta$ in optical analogy, respectively.
Apart from the similarity, we should also point out the difference in the electron effective masses brings new degree of freedom, and as a
result the critical angle, reflection, transmission, and GH shift are more complicated and also interesting. For instance,
when the negative effective electron mass is considered in a semiconductor, e.g. $GaN$ \cite{DragomanJAP},
the electronic analogy of negatively refractive media can be achieved, and the GH shift of ballistic electrons is expected to be negative \cite{Berman}.

\subsection{quantum barrier and well}

\begin{figure}[]
\begin{center}
\scalebox{0.65}[0.65]{\includegraphics{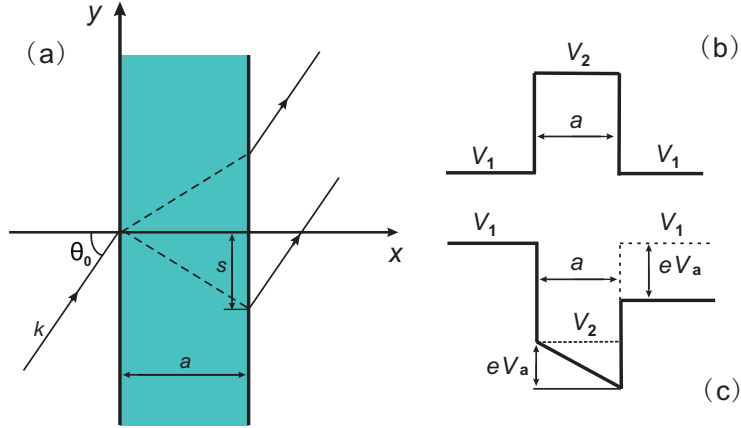}}
\caption{(Color online)  (a) Schematic diagram of (positive and negative) lateral GH shifts
of ballistic electrons propagating obliquely through a quantum barrier or well,
corresponding to a two-dimensional semiconductor potential barrier (b) and well
under external applied electric field (c).} \label{fig.2}
\end{center}
\end{figure}

Based on the previous results \cite{Wilson-G-G,ChenPLA06}, the GH shift in total reflection can be further generalized
to partial transmission in semiconductor quantum barrier and well, as shown in Fig. \ref{fig.2}. First of all, we turn to discuss
the GH shift in transmission through a semiconductor quantum barrier, extending from $0$ to $a$,
where $V_2>V_1$, $V_2$ is the height of
potential barrier, $V_1$ are the potential energies on its two sides, and corresponding electron effective masses are
$m_{2}^{\ast}$ and $m_{1}^{\ast}$. In this case, the critical angle for total reflection is given by Eq. (\ref{critical angle}),
so that the electron wave propagation inside the region of barrier can be divided into two cases: evanescent and propagating waves.
Different from the GH shift in total reflection, the GH shift in transmission though potential barrier
can be enhanced by transmission resonances \cite{ChenPLA06}, where the potential barrier is described in Fig. \ref{fig.2} (b).
Therefore, we just emphasize the propagating case when the incidence angle $\theta_0$ is less than the critical angle $\theta_c$ [see Eq. (\ref{critical angle})] for total reflection.
In this case, the GH shift of the transmitted beam is easily obtained
by stationary phase method
\beq
s= -\frac{\partial \phi}{dk_{y}}|_{\theta=\theta_0},
\eeq
where the phase $\phi$ is given
\begin{equation}
\label{phaseshift}
\tan \phi= \frac{1}{2}\left(
\frac{m^*_2}{m^*_1}\frac{k_x}{k'_x}+
\frac{m^*_1}{m^*_2}\frac{k'_x}{k_x} \right) \tan k'_x a,
\end{equation}
where $k'_x=k'\cos \theta'$, $k'=[2m_{2}^{\ast}(E-V_{2})]^{1/2}/\hbar$, and $\theta'$ is determined by
Snell's law for electron wave, $ \sin\theta'/\sin \theta=
[m^*_1(E-V_1)/m^*_2(E-V_2)]^{1/2}$. Thus, the GH shift in transmission
has the following form:
\begin{equation}
s=\frac{s_{g}}{2 f^2_0}\left[\left(\frac{m_{2}^{\ast}k_{x0}}{m_{1}^{\ast}k'_{x0}}+\frac{m_{1}^{\ast}k'_{x0}}{m_{2}^{\ast}k_{x0}}\right)
-\left(1-\frac{k_{x0}^{'2}}{k_{x0}^{2}}\right)\left(\frac{m_{2}^{\ast}k_{x0}}{m_{1}^{\ast}k'_{x0}}-\frac{m_{1}^{\ast}k'_{x0}}{m_{2}^{\ast}k_{x0}}\right)
\frac{\sin2k_{x0}^{'2}a}{2k_{x0}^{'2}a}\right],
\end{equation}
where $s_{g}=a\tan\theta'_{0}$ is the shift
predicted from Snell's law for electron waves, and $T_{0}=1/f^2_0$ is the
transmission probability, determined by the following complex number,
$$
f e^{i \phi}=\cos k'_x a+
\frac{i}{2}\left(\frac{m^*_2}{m^*_1}\frac{k_x}{k'_x}+
\frac{m^*_1}{m^*_2}\frac{k'_x}{k_x}\right) \sin k'_x a.
$$
In the evanescent case where the incident angle $\theta_0$ is
larger than the critical angle $\theta_{c}$, the above expression is
still valid, only if $k'_{x0}$ is replaced by $i\kappa_{0}$.
Importantly, the GH shift in this case becomes independent of
barrier's thickness, and saturates a constant, which is the order of electron wavelength.
This phenomenon is similar to the optical GH shift and associated with Hartman effect in FTIR \cite{Chen-PRAa,Chen-OL},
Here we are more interested in the GH shift in the propagating case,
instead of evanescent case.

When $k'_{x0}a=m\pi$ $(m=1,2,3...)$, we have $T_{0}=1$, which is so called
transmission resonance. At resonances, and the lateral shift is maximal,
\begin{equation}
s_{max}=s|_{k'_{x0}a=m\pi}=\frac{s_{g}}{2}\left(\frac{m_{2}^{\ast}k_{x0}}{m_{1}^{\ast}k'_{x0}}+\frac{m_{1}^{\ast}k'_{x0}}{m_{2}^{\ast}k_{x0}}\right).
\end{equation}
If the incidence angle $\theta_{0}$ is close to the critical angle
$\theta_{c}$, $k'_{x0}$ is much less than $k_{x0}$, so that
$s_{max}$ is much larger than $s_{g}$. On the other hand, when
$k'_{x0}a=(m+1/2)\pi$, $T_{0}$ reaches its minima and the electron
beam is most reflected back. In this case, the lateral shift is also
minimal,
\begin{equation}
s_{min}=s|_{k'_{x0}a=(m+1/2)\pi}=2s_{g}\left(\frac{m_{2}^{\ast}k_{x0}}{m_{1}^{\ast}k'_{x0}}+\frac{m_{1}^{\ast}k'_{x0}}{m_{2}^{\ast}k_{x0}}\right)^{-1}.
\end{equation}
Obviously, the GH shift is modulated by the
transmission probability $T_{0}$, so that it can be greatly enhanced by transmission resonances
resulting from the multiple reflections inside the barrier.

In analogy with the negative GH shift in dielectric slab \cite{Li-2}, the GH shift of ballistic electrons is found to be
negative as well as positive when the electron beam transmits though semiconductor quantum well, as shown in Fig. \ref{fig.2} (c), acting as quantum slab \cite{ChenJAP09}.
The expression for the GH shift in quantum well is the same as that for quantum barrier. The only difference
considered here is $V_2 < V_1$ for semiconductor quantum well. When the necessary condition
\begin{equation}
\label{inequality}
\frac{m^*_2}{m^*_1}\frac{k_{x0}}{k'_{x0}}+
\frac{m^*_1}{m^*_2}\frac{k'_{x0}}{k_{x0}}<
\left(1-\frac{k'^2_{x0}}{k_{x0}^2} \right)
\left(\frac{m^*_2}{m^*_1}\frac{k_{x0}}{k'_{x0}}-
\frac{m^*_1}{m^*_2}\frac{k'_{x0}}{k_{x0}}\right),
\end{equation}
is fulfilled, the GH shift becomes negative, when the incident angle satisfies
\begin{equation}
\label{restriction to incident angle} \cos \theta_0 <
\left[\frac{m^*_2(E-V_2)/m^*_1(E-V_1)-1}{1+(m^*_2/m^*_1)^2}\right]^{1/2}
\equiv \cos \theta_t.
\end{equation}
Further analysis shows that when the incident energy is in the region of $V_1<E<E_c$, the necessary condition
(\ref{restriction to incident angle}) is satisfied, so that the
lateral shift of ballistic electrons in the quantum slab can be
negative as well as positive. In other word, if the incidence angle $\theta_0$ is larger
than the threshold angle $\theta_t$ the GH shifts in
transmission through a semiconductor quantum slab can be negative for some values $a$, width of semiconductor quantum well,
while the GH shifts are always positive in the semiconductor quantum barrier ($V_2>V_1$).

In addition, when $V_2 < V_1$ is assumed for quantum well,
there exists the critical angle, $\theta_c$, at the electron energies above $E_c$.
In the case of $E>E_c$, the total internal reflection occurs. Thus,
the lateral shifts are always positive, since the necessary
condition (\ref{restriction to incident angle}) is invalid.
On the contrary, when $E>E_c$ for quantum barrier $V_2 > V_1$,
the GH shifts can be negative under some conditions \cite{ChenJAP09}.
All these complicated but interesting behaviors result from the effective masses.
Note that if the effective masses are equal, $m^*_1=m^*_2$, the critical angle for quantum well
does not exist because  the critical energy $E_c$ goes to infinity.

Before we turn to the modulation of GH shifts and its applications, we have to clarify the physical explanation
of positive and negative GH shifts of ballistic electrons in quantum barrier and well. Originally, the GH shift was explained
by the reshaping of totally reflected plane-wave components undergoing different phase shifts, based on Artmann's stationary phase method \cite{Artmann}.
The GH shifts presented here can be further understood from
the destructive and constructive interference due to the multiple reflections and transmission inside the barrier or well.
The negative GH shift is similar to that of light beam transmitted through a dielectric slab \cite{Li-2}, but different from
that in negatively refractive media \cite{Berman,Li-2}. As illustrated in optical analogy,
we have already demonstrated that the negative lateral shift is
produced by the reshaping effects due to the multiple reflections \cite{Chen-JOA}.

\subsection{modulation by external fields}

From the point of view of applications, the controllability of GH shifts by external fields
will provide more flexibility to design the electronic devices, based on
the negative and positive shifts mentioned above. To this end, we first focus
on the control of GH shift by external electric field, taking account into imposing
applied bias field on the semiconductor quantum well, see Fig. \ref{fig.2} (c).
Because the system is
translationally invariant along the $y$ direction, the wave function
in the region of the quantum well under applied electric field can
be expressed as $\Psi (x,y)=\psi(x)e^{ik_y y}$, where the
longitudinal wave packet is determined by,
\begin{equation}
\frac{\partial^2 \psi(x)}{{\partial
x^2}} +\frac{2m^*_2}{\hbar^2}\left(E_x-V_2+
\frac{eV_ax}{a}\right)\psi(x) = 0,
\end{equation}
where $E_x=\hbar^2 k^2_x/2 m^*_2$ is the longitudinal energy and
$V_a$ is the applied biased voltage.
Generally,
the solutions of Schr\"{o}dinger equation for the Hamiltonian with applied biased voltage $V_{a}$ are the well-known linearly
independent Airy functions $Ai(\eta)$ and $Bi(\eta)$, that is
$\psi(x)=M Ai(\eta)+N Bi(\eta)$, where
\beq
\eta=\left(-\frac{2m_{1}^{\ast}eV_{a}}{a\hbar^{2}}\right)^{1/3}\left[\frac{a}{eV_{a}}(E_{x}-V_{2})+x\right].
\eeq
with longitudinal
energy $E_{x}=\hbar^{2}k_{x}^{2}/2m_{2}^{\ast}$. Based on the boundary conditions,
the analytical expressions for transmission coefficient and corresponding phase shift are
obtained, then the GH shift is finally numerically calculated. In Ref. \cite{ChenJAP09}, the numerical results show that the GH shifts with biased voltage
can also be negative as well as positive in the same way as those in absence of external applied electric field, and
the lateral shifts are tuned from negative to positive when the biased voltage $V_a$ is increased.
In addition, we have also illustrated that the parameter $\chi$ of semiconductor has great effect
on the lateral shifts, when we consider the semiconductor quantum well consisting of
$\mbox{Ga}_{1-\chi}\mbox{Al}_{\chi}\mbox{As/GaAs/}\mbox{Ga}_{1-\chi}\mbox{Al}_{\chi}\mbox{As}$.
Thus, this provides a more feasible scheme to control the GH shifts
by using external fields, instead of manipulating the parameters of semiconductor structure, e.g. the width $a$ of quantum well.


\begin{figure}[t]
\begin{center}
\scalebox{0.60}[0.60]{\includegraphics{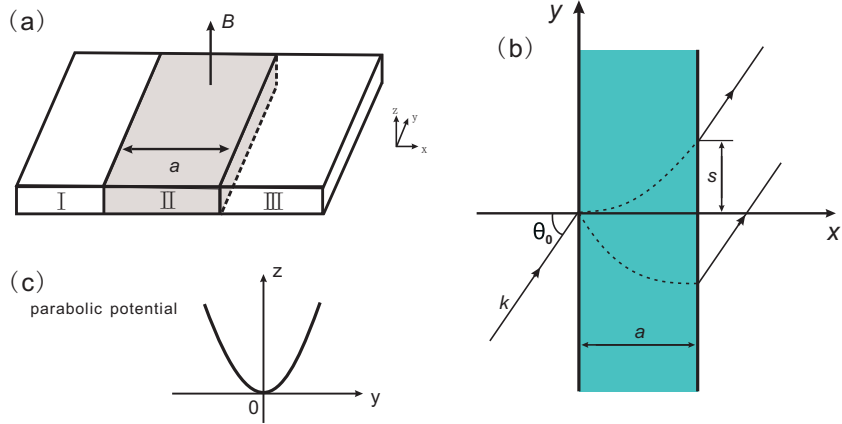}}
\caption{(Color online)  (a) Schematic diagram for 2DEG  with the parabolic quantum well under a uniform magnetic field B along the vertical z-direction.
(b) Negative and positive GH shifts of ballistic electrons are presented in this configuration.
(c) Description of the parabolic confinement potential $H_{conf}$.
}
\label{fig.3}
\end{center}
\end{figure}

Next, we shall discuss that the external magnetic field provide an alternative way to modulate the (negative and positive) GH shifts in quantum well.
As depicted as showin Fig. \ref{fig.3}, we consider the GH shift for ballistic electrons with total energy $E$ in a parabolic quantum well under a uniform magnetic field, in which the
Hamiltonian is given by
\beq
H=\frac{1}{2m^{\ast}}(\textbf{p}-\textbf{A})^{2}+\frac{1}{2}\mu_{B}g\sigma B + H_{conf},
\eeq
with the parabolic confinement energy $H_{conf}=\frac{1}{2}m^{\ast} \omega_{0}^{2}y^{2}$.
This system is particularly interesting, because at high magnetic fields, Landau levels form and quantum Hall effect have no any optical analogy \cite{Datta}.
In such a quantum well, the uniform magnetic field bends continuously the trajectory of electron, exhibiting cyclotron motion,
which implies that the motion of electron has no direct analogy with the linear propagation of light. But the transmitted
electron beam still experiences the GH shift in such quantum systems, and the unique properties of GH shifts depend on not only
incident energy and incidence angle, but also the magnetic field and Landau quantum number. After some lengthy but direct calculations,
the GH shift in quantum well under a uniform magnetic field is obtained, by using the stationary phase method, as \cite{ChenPRB11}
\begin{equation}
\label{Goos-Hanchen shift}
s=\frac{a \tan\theta_0}{2 f^2_0 }\left(1-\frac{k^{'2}_{x0}}{k_{x0}^{2}}\right)\frac{\sin{(2 k'_{x0}a)}}{2k'_{x0
}a},
\end{equation}
where the subscript $0$ means the value at $k_y = k_{y0}$, that is $\theta=\theta_0$, as before, for the plane wave component under consideration
\begin{equation}
f \exp{(i \phi)= \cos{(k'_{x} a)} + i \left(\frac{k^2_{x}+k^{'2}_{x}}{2 k_{x} k'_{x}}\right) \sin{(k'_{x} a})},
\end{equation}
the corresponding longitudinal wave vector $k_{x}=(k^2 -k_{y}^{2})^{1/2}$, and $k'_x$ in the central potential region is
\begin{equation}
\label{wavevectorB}
k'_{x}=\sqrt{\frac{2m^{\ast} \widetilde{\omega}_{c}^{2}[E_{n}-(n+\frac{1}{2})\hbar\widetilde{\omega}_{c}-\frac{1}{2}\mu_{B}g\sigma
B]}{\hbar^{2}\omega_{0}^{2}}},
\end{equation}
with the parameters: $k=(2 m^{\ast} E)^{1/2}/ \hbar$, effective mass of electron $m^{\ast}$, energy $E_n$ ($n$ is Landau level), cyclotron frequency $\omega_{c}=B e/m^{\ast}$,
and $\widetilde{\omega}_{c}^{2}=\omega_{0}^{2}+\omega_{c}^{2}$.
When the only plane wave is considered, the spatial location inside the potential well is around
\begin{equation}
Y \equiv \hbar k'_{x}/eB = \nu (n, k'_x) \frac{\omega_{0}^{2}+\omega_{c}^{2}}{\omega_{c} \omega_{0}^{2} },
\end{equation}
with the velocity
\begin{equation}
\nu (n, k'_x) = \frac{1}{\hbar} \frac{\partial E_n}{\partial k'_x}= \frac{\omega_{0}^{2} }{\widetilde{\omega}_{c}^{2}} \frac{\hbar k'_x}{m^{\ast}} .
\end{equation}
The transverse location for each plane wave eigenstate is proportional to the velocity and magnetic field. From the classical viewpoint,
the spatial displacement can be plausibly explained by Lorentz force \cite{Datta}. As a consequence,
the transverse shifts for the forward and backward propagating states inside the central region are positive and negative,
since Lorentz force is opposite for electrons moving in the opposite direction.
However, the lateral displacement predicted by the GH effect will be totally different.
On one hand, in the propagating case, when the transmission resonances $k'_{x}d = m\pi$
or antiresonances $k'_{x}d = (m+1/2)\pi$ occur, the GH shift is zero, which means the positions in the
$y$ direction are the same for both incident and transmitted
electrons. On the other hand, the GH shifts can be negative and positive, depending not only on the incident energy, but also on magnetic field.
Whereas, in the evanescent case, the GH shifts become always positive, like in semiconductor
barrier. The dependence of GH shift on the strength of magnetic field and Landau energy level provides more freedom to control the GH shifts.
In the propagating case, the GH shifts can be changed from negative to positive by controlling the strength of the magnetic field, and vice versa.
However, the GH shifts finally become positive with increasing the strength of the magnetic field, since the propagation of electrons is actually evanescent.
Based on these phenomena, the spin beam splitter is proposed in \cite{ChenPRB11}, in which the
spin-up and spin-down electron beams can be completely separated by negative and positive GH shifts.
In addition, we can choose the incident energy within the range
of $E^{+}_{c}<E<E^{-}_{c}$, where the critical energy for evanescent waves with spin-up and spin-down polarizations are
\begin{equation}
\label{condition}
E_{c}=(n+\frac{1}{2})\hbar\widetilde{\omega}_{c}+\frac{1}{2}\mu_{B}g\sigma B,
\end{equation}
obtained from Eq. (\ref{wavevectorB}).
This suggests that the spin-down polarized electrons for $E>E^{-}_{c}$ can traverse through the structure
in the propagating mode with high transmission probability, while the spin-up polarized electrons for $E< E^{+}_{c}$ tunnel through it in the evanescent mode with very low
transmission probability. These phenomena describes an alternative physical mechanism to design a spin spatial filter with energy gap, $\Delta E= \mu_B g B$.
In what follows that we will discuss the applications of GH shifts in spintronics.

\subsection{applications in spintronics}

\begin{figure}[t]
\begin{center}
\scalebox{0.55}[0.55]{\includegraphics{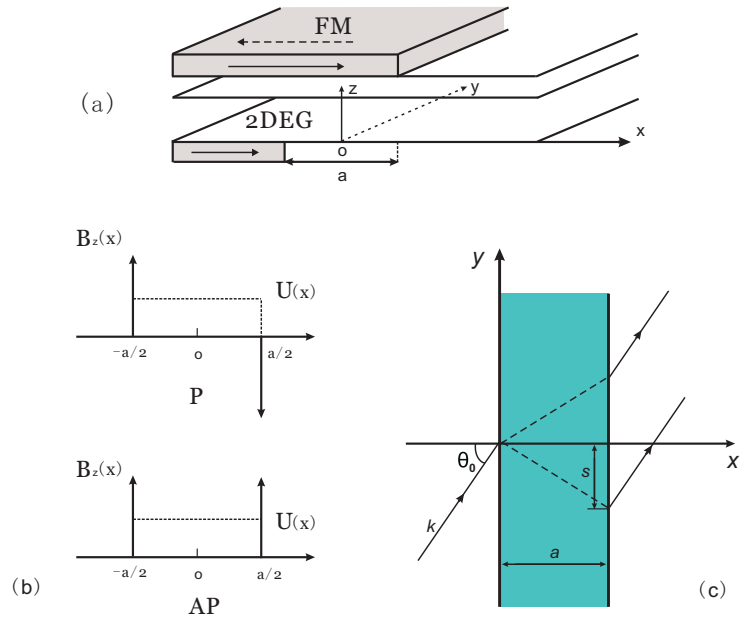}}
\caption{(Color online) (a) Schematic illustration of the magnetic-electric
nanostructure with two metallic FM stripes deposited on top and
bottom of the semiconductor heterostructure; (b) The
magnetic-electric barrier models exploited here corresponds to the P
and AP magnetization configurations of two FM stripes, respectively.
(c) The positive and negative GH shifts in this nanostructure.
}
\label{fig.4}
\end{center}
\end{figure}

Now we shall discuss the spin-dependent GH shift and its application in spintronics. As mentioned before,
the proposal for designing spin beam splitter is that
the spin-up and spin-down polarized electron beams are separated by different GH shifts in parabolic quantum well
under a uniform magnetic field, due to their energy dispersion realtion depending on the spin polarization \cite{ChenPRB11}.
In other realistic electronic devices, a significant class of 2DEG nanostructures, magnetic barriers, can be experimentally realized
by the deposition of a 2DEG in an inhomogenous field, see the references in \cite{ChenPRB08,Lu-PLA,Lu-EPJB,Lu-JMMM,Lu-JAP}. In 2008, we
have proposed for the first time the tunable GH shifts and spin beam splitter in the modulation-doped semiconductor heterostructure,
depositing two metallic FM stripes on top and bottom of the
semiconductor heterostructure \cite{ChenPRB08}. The configuration of such magnetic-electric nanostructure
is shown in Fig. \ref{fig.4}. The Hamiltonian describing
such a system in the $(x, y)$ plane, within the single particle
effective mass approximation, is
\begin{equation}
H=\frac{p_x^2}{2m^*}+\frac{[p_y+e{\bf
A}_y(x)]^2}{2m^*}+U(x)+\frac{eg^*}{2m_0}\frac{\sigma \hbar}{2}
B_{z}(x),
\end{equation}
where $m^*$ is the electron effective  mass and $m_0$ is the free
electron mass, $(p_x, p_y)$ are the components of the electron
momentum, $g^*$ is the effective Land\'{e} factor, $\sigma=+1/-1$ for spin up/down electrons, and ${\bf
A}_y(x)$ is the y-component of the vector potential given, in Landau
gauge, by $\vec{{\bf A}}=[0, {\bf A}_y(x), 0]$. For GaAs, we can choose $B_{z}=0.1$ T,  $g^* = 0.44$, and $m^* = 0.067 m_0$.
Once the Hamiltonian is given, we can follow the similar process to calculate the GH shift in terms of stationary phase method.
It was found that
the GH shifts of transmitted electron beam are negative and positive, and can be controlled by
the magnetic-field strength of ferromagnetic stripes and the applied voltage.  Depending on
the magnetic barriers, described by $B_{z}(x)=B_{1}\delta(x+a/2)-\chi B_{2}\delta(x-a/2)$,
the behaviors of GH shifts for $\chi=+1$ (P case) and $\chi=-1$ (AP case) are totally different with respect to
spin polarization. In the AP configuration, the GH shift
depends on the electron spin due to its asymmetry, so
that the spin-polarized electron beam can be separated spatially,
while in P configuration GH shift does not depend on spin polarization.
As a result, the negative and positive GH shifts depend on electron spins
when two $\delta$-magnetic barriers point at the same direction, thus
a spin beam splitter that separates spin-up and spin-down electron beams
by their corresponding GH shifts can be proposed in parallel double $\delta$-magnetic barrier nanostuctures.
The proposed spatially separating spin filter and spin beam  splitter are
different from those designed by the refraction of a spin electron beam at
the interface by considering spin-orbit coupling of Rashba and Dresselhaus types \cite{Khodas,Zhang}.

Following the work in \cite{ChenPRB08}, Lu's group has also
investigated the tunable GH shift and spin beam splitter in the
magnetic barrier nanostructure and its variants
\cite{Lu-PLA,Lu-EPJB,Lu-JMMM,Lu-JAP}. For example, the realistic
magnetic barriers created by lithographic patterning of
ferromagnetic or superconducting film have been further considered
\cite{Lu-JAP}. Due to intrinsic symmetry, only nanostructures with
symmetric magnetic field possess spin-dependent GH shifts. The
control of GH shifts over the incident angle, incident energy, size
and position of ferromagnetic stripe is of great benefit to design a new
type of spin filter, spin injection, and spin beam splitter in
such magnetic-electric nanostuctures.

\section{Electronic Goos-H\"{a}nchen shifts in graphene}

\label{graphene}

\subsection{single interface}

\begin{figure}[]
\begin{center}
\scalebox{0.7}[0.7]{\includegraphics{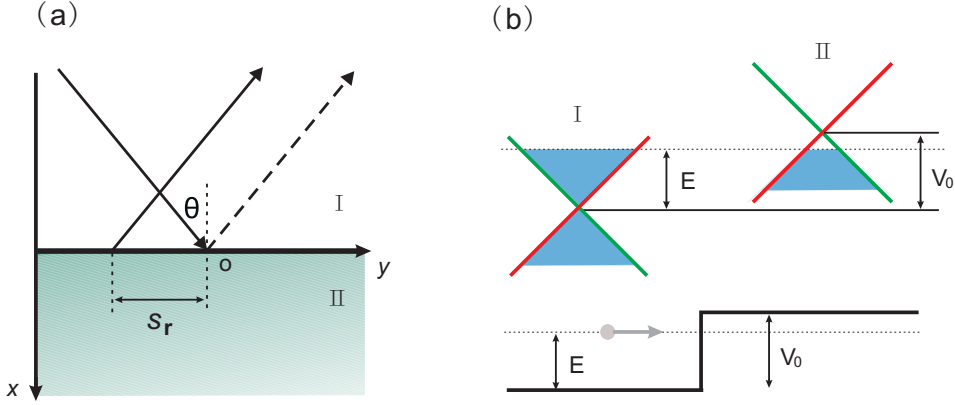}}
\caption{(Color online)  (a) Schematic diagram of total internal reflection and negative GH shift at a sharp $p$-$n$ junction.
(b) Low-energy linear spectrum and potential diagram.}
\label{fig.5}
\end{center}
\end{figure}

In this section, we shall turn to the GH shift in graphene, which is different from that in semiconductor.
As we know, graphene is a two-dimensional zero-gap
semiconductor with linear dispersion relation, $E = \hbar
k \upsilon_{F}$, thus the low-energy quasi-particles are formally
described by the Dirac-like Hamiltonian \cite{Castro},
$\hat{H_{0}}=-i\hbar v_{F}\sigma \cdot \nabla $, $v_F \approx 10^{6} m
\cdot s^{-1}$ is the Fermi velocity, $k$ is the Fermi wave vector,
and $\sigma=(\sigma_{x}, \sigma_{y}) $ are the Pauli matrices. The linear energy dispersion and two spinor component of
the wave function differentiate the electron wave propagation in graphene
from that in semiconductor, which leads to many interesting optic-like phenomena, such as
negative refraction \cite{Cheianov}, collimation \cite{Park},
Fabry-Perot interference \cite{Shytov}, Bragg reflection \cite{Ghosh} and waveguide \cite{Williams}.
To consider the GH shift in graphene, the incident electron beam is assumed to be
\begin{equation}
\Psi_{in}=\int_{-\infty}^{\infty}dk_y  A(k_y-k_{y0}) e^{i (k_x x + k_y y)}
\left(
\begin{array}{c}
e^{-i\theta/2} \\
e^{i\theta/2} \\
\end{array}
\right),
\end{equation}
and the reflected one is thus expressed as
\begin{equation}
\Psi_{r}=\int_{-\infty}^{\infty}dk_y r(k_y) A(k_y-k_{y0})e^{i (- k_x x + k_y y)}
\left(
\begin{array}{c}
-ie^{i\theta/2} \\
ie^{-i\theta/2} \\
\end{array}
\right),
\end{equation}
where $A(k_y-k_{y0})$ is the angular spectral distribution around the central wave vector $k_{y0}$ and the reflection
amplitude is $r(k_y)= |r|\exp{[i \phi_r (k_y)]}$. The two upper and lower
components $\Psi_{in,\pm}$ at the interface $x=0$ centered at two
different mean $y$ coordinates, $\bar{y}_{\pm}^{in}=\pm\frac{1}{2}\theta'(k_{y0})$.
The two components $\Psi_{r,\pm}$ at the interface $x=0$ centered at
$\bar{y}_{\pm}^{r}=-\phi'_r (k_{y0})\mp\frac{1}{2}\theta'(k_{y0})$.
So the upper spinor component displaced along the interface
by an amount
$\sigma_{+}=\bar{y}_{+}^{r}-\bar{y}_{+}^{in}=-\phi'_r (k_{y0})-\theta'(k_{y0})$,
while the displacement of lower spinor component is given by
$\sigma_{-}=\bar{y}_{-}^{r}-\bar{y}_{-}^{in}=-\phi'_r (k_{y0})+\theta'(k_{y0})$.
Different from that in semiconductor, the GH shift in graphene is
defined as
\begin{equation}
\label{sgraphene}
s_r=\frac{1}{2}(\sigma_{+}+\sigma_{-})=- \frac{\partial \phi_r}{\partial k_y}|_{k_y=k_{y0}},
\end{equation}
which means the average displacement of upper and lower spinor components. Noting that different
spin component for incident wave will lead to different definition of GH shift in graphene.
In order to clarify the difference from the GH shift in semiconductor, we would like to repeat
the calculations of GH shift in the literatures \cite{Beenakker-PRL,Zhao}.
As shown in Fig. \ref{fig.5}, we consider the incident electron wave with incidence angle $\theta$. Based on the Dirac-like equation described above, we
can write down the wave functions in the two different regions of I and II.
By defining $k=|E|/\hbar v_{F}$ and $k'=|E-V_{0}|/\hbar v_{F}$, we have
\begin{eqnarray}
\psi_{I} =  \frac{1}{\sqrt{2}} \left(
                      \begin{array}{c}
                       e^{-i\theta/2} \\s e^{i\theta/2}
                      \end{array}
                    \right)e^{i k_{x} x + i k_{y} y}
           +\frac{r}{\sqrt{2}} \left(
                      \begin{array}{c}
                        - i e^{i\theta/2} \\s i e^{-i \theta/2}
                      \end{array}
                    \right) e^{-i k_{x} x + i k_{y} y},
\end{eqnarray}
with $k_{x}=k \cos\theta$, $k_{y}=k \sin\theta$, and
$s=\textrm{sgn}(E)$, where $r$ is the reflection amplitude. In
region II,  the transmitted wave is given by
\beqa \psi_{II}=
\frac{t}{\sqrt{2}}  \left(
                     \begin{array}{c}
                        e^{-i\varphi/2} \\s' e^{i\varphi/2}
                      \end{array}
                    \right) e^{i k'_x x + i k_{y} y},
\eeqa
with $k'_{x}=k' \cos\varphi$ and $s' =\textrm{sgn}(E-V_{0})$. By using boundary conditions at $x=0$,
we can calculate the coefficients $r$ and $t$, where $r$ is given by
\begin{equation}
r =
\frac{i(e^{i \theta}-ss' e^{i \varphi})}{
1 + ss' e^{i \theta} e^{i\varphi_r}}.
\end{equation}
When the incidence angle $\theta$ is larger than critical angle $\theta_c =\arcsin|V_{0}/E-1|$,
the total reflection occurs. In this case, $k'_{x}= i \kappa = i (k_y^2 -k'^2)^{1/2}$, the reflection coefficient
becomes
\begin{equation}
r= |r|\exp{[i \phi_r (k_y)]} = \frac{i e^{i \theta} ss' k' + (\kappa +k_y)}{ss' k' + i e^{i \theta}(\kappa +k_y)}.
\label{r}
\end{equation}
According to the above definition (\ref{sgraphene}), the GH shift at the single interface, as shown in Fig. \ref{fig.5}, is obtained as \cite{Beenakker-PRL}
\begin{equation}
s_r =\frac{1}{\kappa_0} \frac{k_{y0}^2 + ss' k^2 (|V_0/E -1|)}{k_{x0} k_{y0}}.
\end{equation}
Noting that $0$ in the subscript means the value at $k_y =k_{y0}$ and the GH shift in total reflection
is again the order of electron wavelength, $1/\kappa_0$.
In general, when graphene $n$-$n$ interface (when $E>V_{0}$) is considered,
$ss'=1$, thus the GH shift is always positive, like the GH shift at the single semiconductor interface.
Interestingly, the negative GH shift appears at a graphene $p$-$n$
interface (when $E<V_{0}$), due to $s s'=-1$, for the incidence angles satisfying
\beq
\theta_{c}<\theta_0 <\theta^{\ast}=\arcsin\sqrt{\sin\theta_{c}}.
\eeq
On the contrary, when $\theta_0 >\theta^{\ast}$,  the GH shift becomes positive, regardless of the relative magnitude of $E$ and
$V_{0}$.

To understand the positive and negative GH shifts in graphene, we may define the
effective refractive index as $n_{e} = E/ (E-V_0)$. In the case of $E<V_0$, the Klein tunneling is analogous to
the phenomenon of negative refraction in metamaterial with $n_{e} <0$. On the contrary, the classical motion for
$E>V_0$ corresponds to the positive refractive index ($n_e >0$) in the normal dielectric. Since the link between Klein
tunneling and negative refraction, many intriguing phenomena in metamaterial, for example, negative refraction, Veselago lens and focusing \cite{Cheianov},
have been transferred to the electron transports in graphene. In addition, the optical analogy of
graphene has been also proposed in photonic crystals \cite{Haldane,Peleg,Sepkhanov} and negative-zero-positive index metamaterial (NZPIM) \cite{WangOL,WangEPL}.
For more information, we refer the readers to the recent review in \cite{Bookmy}.


\subsection{single, double and multiple barriers}


Recent publications have shown that the GH shifts in graphene can be greatly enhanced by transmission resonances
in various graphene-based nanostructures including single \cite{Chen-EPJB}, double \cite{Guo} and multiple barriers \cite{Chenarxiv}.

In graphene-based barrier, the expression of GH shift are similar to that in semiconductor barrier, so we will not repeat them.
But since the unique properties of electron transport in graphene, the GH shifts are simply discussed
in two difference cases: Klein tunneling and classical motion \cite{ChenAPL}.
In classical motion, the critical angle becomes
\begin{equation}
\label{critical angle-1}
\theta_{c}=\arcsin\left(1-\frac{V_{0}}{E}\right),
\end{equation}
when $E>V_{0}$ is satisfied. Thus, the electron wave propagations can be propagating
and evanescent cases, which depend on the incidence angle below or above the critical angle for total reflection.
In both cases, the GH shifts are always positive as those in the 2D
semiconductor barrier \cite{ChenPLA06}.
In Klein tunneling case, there is a critical angle for total reflection, defined as
\begin{equation}
\label{critical angle-2}
\theta_{c}=\arcsin\left(\frac{V_{0}}{E}-1\right),
\end{equation}
when the condition $E<V_{0}<2E$ is satisfied. When the incidence angle $\theta_0$ is less than the critical angles $\theta_c$, that is, $\theta_0<
\theta_{c}$,  the GH shifts can be positive
as well as negative. Particularly, the lateral shifts can be
positive for a thin barrier, while at some resonance or
anti-resonance points they become negative. And the negative lateral
shifts can be enhanced by the transmission resonances.
When $\theta_0 > \theta_{c}$, the GH shift in the
evanescent case is about the order of electron wavelength, and
becomes independent of barrier's thickness. More interestingly, the
saturated GH shift is negative when the incidence angle satisfies
$\theta_{c}<\theta_0<\theta^{\ast}$, while it becomes positive when $\theta_0>\theta^{\ast}$.
This result is related to pseudospin degree, as mentioned in \cite{Beenakker-PRL}.
As a consequence, the sign change of the GH shifts appears at different incidence
angles, which also provides the freedom to control the GH shift.
As a matter of fact, the positive and negative GH shifts discussed here
remind us those of light beam transmitted through a slab of
negative-index metamaterial \cite{Li-2}. In the slab of negative-index metamaterial,
the negative index contributes to the negative lateral shift, while
the positive lateral shift does result from the reshaping effect due to the multiple reflections.
This provides the physical explanation of positive and negative GH shifts in the graphene barrier,
which is totally different from those in semiconductor quantum slab \cite{ChenJAP09}.
In semiconductor quantum slab, the negative lateral shift has nothing to do with Klein tunneling (negative refraction in the language of optics),
and does originate from the beam reshaping resulting from multiple reflections inside the slab \cite{Li-2}.
Our recent work is devoted to the optical simulation of transmission gap and Klein tunneling in graphene barrier \cite{Chen-PRAb},
by using the negative-zero-positive index metamaterial (NZPIM) \cite{WangOL,WangEPL}, in which
the similar negative and positive GH shifts, relevant to Klein tunneling and classical motion, are also found.

Most recently, the GH shifts in graphene-based barrier are directly
generalized to graphene double barrier and multiple barrier (superlattice)
structures. Different from the GH shifts in graphene barrier, the GH shifts
in double barrier structure can be greatly enhanced \cite{Guo} inside the
transmission gap in a single barrier \cite{Chen-EPJB,ChenAPL}. This giant GH
shift is attributed to the quasibound states formed in double
barriers. Moreover, they have made the conclusion
that the smoothness of the potential barrier will not restrict the application of the giant GH shifts,
by studying a realistic potential which varies smoothly on the scale of the graphene lattice constant.
(We should point out that the similar analysis of electronic GH shifts is also interesting for the smooth interfaces 
in semiconductor.)
In subsequent work, we have also investigated the giant negative
and positive lateral shifts for the transmitted electron beam
through monolayer graphene superlattices \cite{Chenarxiv}. The GH
shifts, depending on the location of new Dirac point, can be
negative as well as positive. Especially when the condition $q_A
d_A= -q_B d_B= m \pi$ ($m=1,2,3...$) is satisfied, the lateral
shifts can be controlled from negative (positive) to positive
(negative) when the incident energy is above (below) the Dirac
point, by increasing the incidence angle. In addition, the lateral
shifts can be greatly enhanced by the effect of defect mode inside
the zero-$\bar{k}$ bandgap. Actually, the optical analogy of giant GH shifts
in graphene superlattice can be realized near the band-crossing structure of one-dimensional photonic crystals containing left-handed metamaterials \cite{WangAPB}.
In a word, the tunable and giant GH
shifts will have potential applications in the graphene-based
electron wave devices, such as electronic wave filter and beam
splitter.

\subsection{strain effect}


As far as we know, the GH shifts in graphene can be tuned by external electric and
magnetic fields. For example, Sharma and Ghosh \cite{Manish} have
studied the electronic GH shifts in terms of the analogy between
light propagation and electron transport in graphene. The GH shifts
are shown to be controlled by adjusting electric and magnetic barriers.
In recent years, more attention has been paid on mechanical strain,
which induces a pesudo-vector potential, and can be used to manipulate electron
transport without using external fields. Motivated by these
progresses, Chang \textit{et. al.} \cite{Wu} have demonstrated that the
electrons in opposite valleys ($K$ and $K'$) show different
Brewster-like angles and GH shifts, by engineering of local strains.
In the strained graphene, the low-energy electrons can be well
described by the effective Hamiltonian $H=v_{F}\sigma^{i}\cdot(p+\xi
A^{i}/v_{F})+V^{i}$, where the superscript $i$ indicates the
different regions, $V^{i}$ is the electrostatic potential in the
region $i$, $\xi=\mp1$ labels $K$ and $K'$ valleys, and the Landau gauge ${\bf A}= (0, A_y, 0)$ is used here.
Here we focus on the analogy between the strained graphene and metamaterial.
For simplicity, $V^{i} =0$ can be set, since the valley-dependent Brewster
angle is the gauge vector $\xi A$. Let us start by considering electron
with $k = E/ \hbar v_F$ reflection from a region of uniform uniaxial strain.
In the strained region, we have $k^2_{x, \xi} + (k_y+\xi A_y)^2 = k^2$, so that
the critical angle for total reflection is
\beq
\theta_{c, \xi} = \arcsin{\left(\frac{k_y +\xi A_y}{k}\right)},
\eeq
and the effective refractive index in optics can be defined as
$
n_{e,\xi} = (k_y +\xi A_y)/ k_y
$.
Note that when $k_y (k_y +\xi A_y) <0 $, the refractive index of the strained graphene is negative just like for a metameterial with negative refractive index.
As pointed out in Ref. \cite{Wu}, the refractive index can be tuned  mechanically in a large range, which is not so for the metamaterial in optics.
Besides, the perfect transmission through a region of uniform uniaxial strain with width $D$ has different windows in the $K$ or $K'$ valley, since
the above critical angle has a different dependence on the vector potential for two valleys denoted by $\xi = \pm 1$. For some incidence angles,
we can achieve totally transmitted electron beams for one valley and totally reflected electron beams for the other valley simultaneously. These characteristic angles are further analogous
to the so-called Brewster angle in optics. Going back to the case of single interface, we apply stationary phase method, and calculate the GH shift for
a uniform uniaxial strain in graphene \cite{Wu}
\begin{equation}
s_r = \frac{2k_{y0}+\xi A_{y}}{k_{x0}\kappa_{0 \xi}},
\end{equation}
from the reflection coefficient
\begin{equation}
r =\frac{k_{x}+i(k_{y}-k'_{y}-\kappa)}{k_{x}-i(k_{y}-k'_{y}-\kappa)}=e^{i 2\phi_r (k_y)},
\end{equation}
and phase shift
\begin{equation}
\phi_r (k_y) =\arctan{\left(\frac{k_{y}-k'_{y}-\kappa}{k_{x}}\right)},
\end{equation}
with $k'_y = k_y +\xi A_y$ and $\kappa = (k'^2_{y} - k^2)^{1/2}$ for the total reflection when incidence angle above the critical angle.
Again the subscript $0$ means the value at $\theta= \theta_0$, where $\theta_0$ is the incidence angle of electron beam,
and $\theta$ is the incidence angle of plane wave under consideration.
Clearly, the GH shifts can be positive or
negative, and the strain effect will provide the totally different scheme to control the GH shift in graphene,
which is different from the control of GH shifts by external electric \cite{Beenakker-PRL} and/or
inhomogeneous magnetic field \cite{ChenPRB11,ChenPRB08}. This
valley-dependent GH shift will further give rise to the application in valleytronics \cite{Zhai,CaoPhysB}.
Exhibiting a close analogy with optical GH shift in dielectric slab \cite{Li-2}, the GH shift
of the transmitted electron beam for a normal/strained/nomal graphene junction
can be further remarkably enhanced by the valley-dependent transmission resonances, which
has been already utilized to design a valley beam splitter in \cite{Zhai}.

Finally, we shall emphasize that the strain effect really provides a useful way to
tune GH shifts in graphene, as compared to the controllability in semiconductor.
The two valleys in graphene are related by time reversal symmetry and act in much the same way
as electron spin in spintronics. So valley-dependent GH shift and its modulation in graphene might be
useful in graphene-based devices, referred to as valleytronics. Thus this research line is promising and
still worthwhile explored. One can further consider the valley-dependent GH shifts in (gapped) graphene
with spin-orbit coupling and bilayer graphene, as the extension in \cite{Chen-EPJB,Cheng}.

\section{Summary and outlook}
\label{summary}

We have reviewed the electronic GH shifts in semiconductor and
graphene-based nanostructures, including single, double barriers and
superlattices. The common feature is that the GH shift can be
enhanced by transmission resonances, and the (negative or positive) lateral shifts can
be controlled by various approaches. From the point of view of application, the
controllability of GH shifts can be realized by using external
electric/magnetic fields in semiconductor and even by adjusting
strain in graphene. The spin (or valley)-dependent GH shifts also
open the new possibility to design the electronic devices, such as
spin (or valley) beam splitters and filters.

Usually the analogies between phenomena occurring in two different physical systems open
a route to find new effects or tow translate techniques or devices, and quite often help to understand
both systems better. In this emerging field,
we have gained an increased understanding and intriguing phenomena,
and new physical effects of electronic GH shifts. However, there still exist several open
questions. For the application, the measurement of electronic GH shifts for
electron beams in semiconductor or graphene remains challenging with several possible reasons,
for example, preparation of required (well-collimated) electron beams, electron scattering, the smoothness of potential,
and the smallness of GH shifts. These imply the electronic systems are different and more complicated, which sets the limits of the optical
analogy in solid state. For example, in the recent experiment on electron waveguide in graphene \cite{Williams},
the GH shift is about $1$ nm ($\simeq 2 \pi/k$) under typical doping conditions.
With the estimated disorder to be of the same order of few nanometer, hence it is impossible to observe
the signature of GH effect, i.e. a quantized conductance of $8e^2/h$ at guide thickness of
$1.5$ nm \cite{Beenakker-PRL}. Theoretically, the electronic GH shift may have the close connection with the electron transport
in semiconductor and graphene, since there is a hint that the lateral and angular GH shifts are analogous to the
skew scattering and side jump \cite{Dyakonov}, but without pursuing (see Fig. 8.4 in reference \cite{Dyakonov}).
Therefore, the spin-dependent GH shifts are considered to be
relevant to anomalous Hall effect \cite{Sinitsyn,Bliokh}. But so far the spin-dependent GH shifts presented in this review
have nothing to do with the spin-orbit coupling. So the spin-dependent GH shift and relevant electronic
transport in presence of spin-orbit coupling  might
be extremely interesting and significant for further work, and will definitely
help to understand the beam shifts and spin-orbit coupling in optics.

Last but not least, recent theoretical prediction \cite{Bliokh-PRL} and experimental
realization \cite{Uchida,Verbeeck} on (GH or IF) beam shifts,
electron vortex beams carrying orbital angular momentum
pave new avenues to investigate angular momentum, spin-orbit interaction \cite{Bliokh-2}, and
relativistic) analogy of spin Hall effect \cite{Bliokh-3}, which are beyond the scope of this review but worth mentioning.

\ack{
The work
was supported by the National Natural Science
Foundation of China (Grant Nos. 60806041, and 61176118), and Shanghai Rising-Star Program (Grant Nos. 08QA14030 and 12QH1400800).
Y. B. also thanks the funding by Basque Government (Grant No. BFI-2010-255).
}

\end{document}